**Irene S. Gabashvili**

Aurametrix, USA

https://aurametrix.com

# Calibrating Resident Surveys with Operational Data in Community Planning

## Abstract

Community associations rely heavily on resident surveys to guide decisions about amenities, infrastructure, and services. However, survey responses reflect perceptions that may not directly correspond to underlying operational conditions. This study bridges that gap by calibrating survey-based satisfaction measures against objective utilization data.

Using parking and facility data from Tellico Village, we map perceived problem rates to utilization exceedance probabilities to estimate behavioral congestion thresholds. Results show that dissatisfaction emerges near effective capacity - once spatial, temporal, and informational constraints are considered - rather than at nominal capacity limits. Perceived difficulty is concentrated among active users and is shaped by operational frictions and incomplete system knowledge.

These findings demonstrate that perceived congestion reflects constraints on access and reliability, not simply physical shortages. By distinguishing between effective and nominal capacity, the proposed framework enables more accurate diagnosis of system conditions. We propose incorporating behavioral metrics into community performance frameworks to support better decision-making, reduce unnecessary capital expansion, and target operational improvements more effectively.

# Introduction

Community associations and master-planned developments rely extensively on resident surveys to guide decisions regarding amenities, infrastructure, and service delivery. These surveys play a central role in governance, informing capital investment, operational changes, and long-term planning. In active lifestyle communities such as Tellico Village, where shared facilities are integral to daily experience, survey-based satisfaction measures often serve as primary indicators of perceived service quality and reliability and community wellbeing.

However, surveys capture perceptions, not direct measures of operational conditions. A persistent challenge is the disconnect between reported experience and actual system performance. Satisfaction is shaped by expectations and perceived discrepancies rather than objective conditions, leading to systematic biases in self-reported data [3]. Users also misperceive conditions, overemphasizing visible issues and relying on heuristics such as accessibility and order [4]. As a result, survey responses can misrepresent underlying system performance when interpreted in isolation.

Evidence from parking systems illustrates this gap. Behavior is strongly influenced by perceived convenience, including preferences for locations that minimize walking distance and uncertainty [5], while spatial imbalances - high demand near entrances alongside underused capacity elsewhere - drive dissatisfaction even when supply is sufficient [6]. These findings suggest that users respond to effective, usable capacity rather than nominal supply, and that interventions aligning system design with user perception are often more effective than purely informational approaches [3]. At the same time, advances in measurement technologies have made it possible to observe infrastructure utilization continuously and at high resolution. Parking occupancy, facility usage, and scheduling activity can now be tracked over time, enabling direct comparison between perceived and observed system performance [3]. Yet these objective data streams are typically analyzed separately from survey responses, leaving decision-makers with two incomplete perspectives: what people experience and what systems actually do. Bridging this gap is essential for accurate diagnosis of performance issues and to avoid misallocation of resources.

This paper links survey responses to objective operational data, treating them as complementary signals of an underlying behavioral process. Survey responses are interpreted probabilistically, where reported problems reflect the likelihood that system conditions exceed a behavioral tolerance threshold.

Using Tellico Village as a case study, we combine parking surveys with continuous utilization measurements and facility-use data [7]. We introduce a calibration framework that maps perceived problem rates to utilization exceedance probabilities, yielding behavioral tolerance thresholds. This approach builds on prior work that modeled resident satisfaction as a function of service quality and cost using a utility-based formulation [8], extending it to incorporate observed system performance and behavioral responses to service reliability. This paper contributes a quantitative calibration framework that maps survey-based perceived problem rates to empirical utilization distributions, enabling direct estimation of behavioral congestion thresholds and effective capacity.

## Methods

We develop a unified framework in which survey responses are interpreted as probabilistic measures of perceived congestion (PPR), while operational data characterize the distribution of actual system utilization. Calibration links these components through behavioral tolerance thresholds, enabling a direct mapping between perceived and observed conditions. This integrated approach allows us to estimate effective capacity and to identify constraints driven by spatial, temporal, and informational factors, rather than by physical supply alone.

Portions of the survey data have been reported in prior work [3,7] (e.g., parking, restaurants, and comprehensive surveys), while additional datasets are introduced here for the first time (e.g., meetings). Similarly, some operational measurements - such as parking utilization - were previously published [3], whereas others are newly analyzed in this study.

### Data Sources

We combine resident survey data with continuous operational measurements from shared facilities in Tellico Village.

Facilities are abbreviated as follows: YC (Yacht Club), WC (Wellness Center), CRC (Chota Recreation Center), TGB (Toqua Golf Clubhouse/Bar & Grill), KCC (Kahite Clubhouse) and TB (Tugaloo Beach) [3]. Neighborhoods are designated by letters A–H [7].

Survey responses provide, for each facility $f$ and neighborhood $n$, the number of visits and the number of visits during which residents reported access or parking problems.

Operational data consist of time-resolved measurements of facility utilization: for parking, utilization is derived from continuous vehicle counts; for meeting spaces, it is constructed from time-stamped booking records.

## Perceived Problem Rate (PPR)

For each facility-neighborhood pair, the **Perceived Problem Rate (PPR)** is defined as:

$$PPR_{f,n} = \frac{\text{Problems}_{f,n}}{\text{Visits}_{f,n}}$$

PPR represents the empirical probability that a resident experiences a problem conditional on visiting a facility.

## Utilization and Survival Function

Let **utilization** at facility $f$ at time $t$ be:

$$U_f(t) = \frac{\text{Observed Occupancy}_f(t)}{\text{Capacity}_f}$$

To characterize utilization dynamics, we construct the empirical survival function:

$$S_f(u) = P(U_f(t) \geq u)$$

This represents the fraction of time that utilization exceeds level $u$.

The key calibration equation in this study is:

$$PPR_f = P(U_f(t) \geq u^*)$$

were

- $U_f(t)$: utilization at facility $f$ at time $t$
- $S_f(u) = P(U_f(t) \geq u)$: survival function
- $u^*$: behavioral tolerance threshold
- $PPR_f$: perceived problem rate

## Behavioral Tolerance Threshold

We interpret survey responses probabilistically by linking perceived problems to utilization exceedance. The **behavioral tolerance threshold** $u^*_{f,n}$ is defined implicitly as:

$$S_f(u_{f,n}^*) \approx PPR_{f,n}$$

Equivalently,

$$u_{f,n}^* = S_f^{-1}(PPR_{f,n})$$

This formulation treats reported problems as the probability that system conditions exceed a resident-specific tolerance level. It provides a direct mapping between subjective perception and observed system performance.

## Statistical Testing

To evaluate whether perceived problems vary systematically across neighborhoods, we perform chi-square tests of independence using counts of problems and non-problems (Visits - Problems).

$$H_0: \text{Problems are independent of neighborhood}$$
$$H_1: \text{Problems depend on neighborhood}$$

## Meeting Space Utilization

To maintain consistency with parking, meeting-space utilization is defined as the proportion of schedulable time in use:

$$U_f(t) = \frac{\text{Booked Time}_f(t)}{\text{Total Schedulable Time}_f(t)}$$

For facilities with multiple rooms:

$$U_f(t) = \frac{\sum_{r=1}^{R_f} \text{Booked Time}_{f,r}(t)}{\sum_{r=1}^{R_f} \text{Available Time}_{f,r}(t)}$$

This time-based formulation allows direct comparison with parking occupancy and enables estimation of utilization distributions and corresponding behavioral thresholds.

## Distance and Facility Choice

All Tellico Village addresses, previously compiled as described in [9], were geocoded using *Geocode by Awesome*. Each address was classified as either a developed residence or an undeveloped lot based on the property database. For each neighborhood, average geographic coordinates were computed across all residential addresses to obtain a representative centroid. These neighborhood centroids were then used to calculate average distances to each facility. This approach provides a consistent measure of spatial separation between neighborhoods and facilities while smoothing parcel-level variation and enabling comparison with neighborhood-level survey responses.

To analyze spatial effects, we model visit frequency as a function of distance between neighborhoods and facilities. For each neighborhood–facility pair:

- $V_{n,f}$: average visit frequency
- $d_{n,f}$: distance between neighborhood centroid and facility

We assume an exponential distance-decay relationship:

$$V_{n,f} = \alpha_f e^{-\beta_f d_{n,f}}$$

This model is estimated separately for each facility to allow heterogeneous decay rates.

### Central Access Index

To capture overall integration into the facility network, we define a **Central Access** measure:

$$CA_n = \sum_f W_f \cdot e^{-\beta_f d_{n,f}}$$

where:

- $W_f$: average visit frequency across neighborhoods (facility importance)
- $\beta_f$: distance sensitivity

Total neighborhood engagement is then:

$$V_n = \sum_f V_{n,f}$$

This distinguishes facility-specific proximity effects from broader network accessibility.

## Utility-Based Satisfaction Framework

Prior work [8] modeled resident satisfaction as a function of quality and cost using a utility-based formulation:

$$U(c, q) = W_c \cdot U(c) + W_q \cdot U(q)$$

where:

- $U(q)$ represents utility derived from service quality
- $U(c)$ represents disutility associated with cost
- $W_c$ and $W_q$ are weighting parameters reflecting relative importance

This framework captures trade-offs between service improvements and financial burden but does not explicitly account for behavioral responses to service reliability or uncertainty.

## Prospect Theory Extension

To incorporate behavioral effects, we extend the utility framework using reference-dependent preferences from Prospect Theory. Let $r$ denote the resident's expectation (reference point) and $x$ the realized service outcome.

The value function is defined as:

$$V(x - r) = \begin{cases} (x-r)^\alpha, & \text{if } x \geq r \\ -\lambda(r - x)^\beta, & \text{if } x < r \end{cases}$$

where:

- $\lambda > 1$ represents loss aversion (losses weigh more heavily than gains)
- $0 < \alpha, \beta < 1$ represent diminishing sensitivity

Under this formulation, negative deviations from expectations (e.g., inability to find parking or secure a meeting room) have a disproportionately large impact on perceived satisfaction. This provides a behavioral explanation for why dissatisfaction emerges sharply near capacity thresholds, where the risk of service failure increases, even if average availability remains acceptable.

## Results

Figure 1 shows that both visits and parking problems vary systematically with distance, but not uniformly across facilities. Central destinations (e.g., YC and WC) exhibit strong attraction effects, while more local facilities follow clearer distance decay. Results are consistent across independent surveys, with strong alignment in visit frequencies (e.g., correlations up to r≈1.0), indicating that patterns reflect real behavior rather than survey noise.

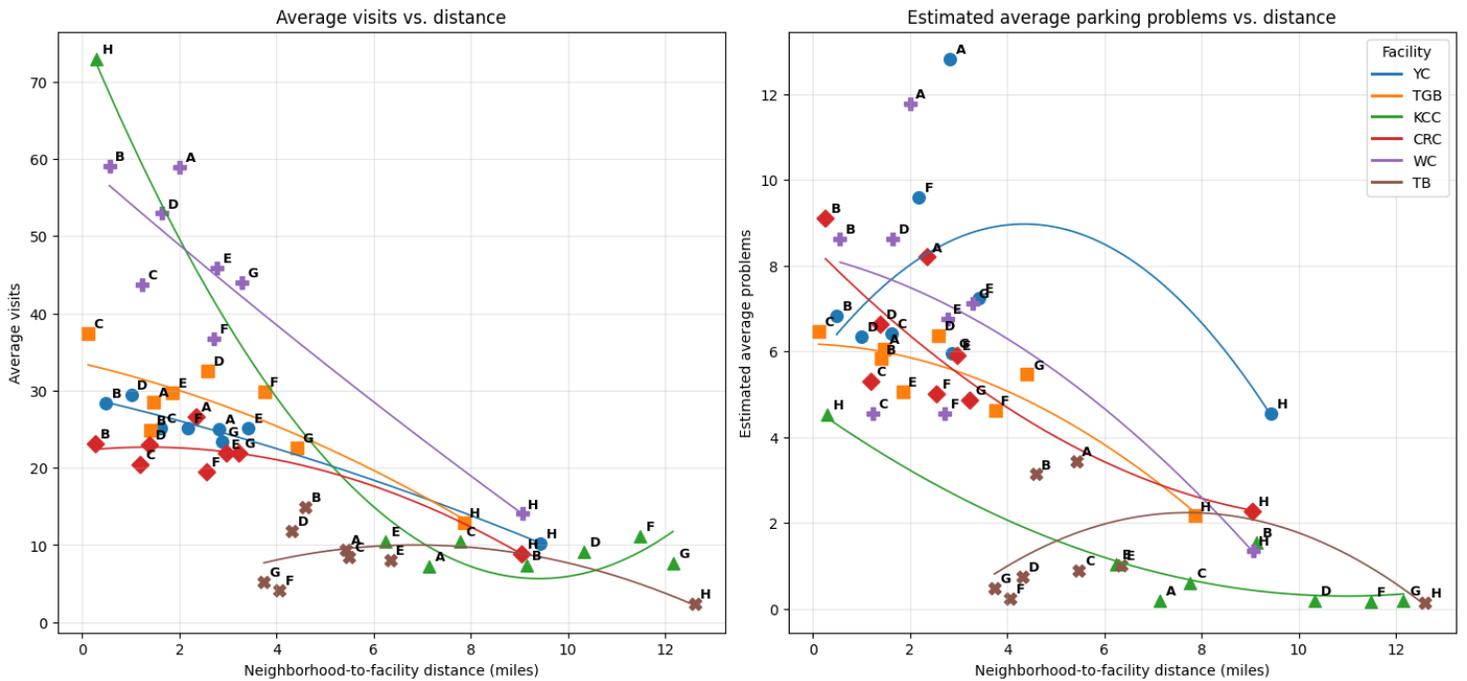

Figure 1 Estimated average visits and parking problems vs. distance. Neighborhoods are designated by letters A–H [7], and facilities are color-coded.

Restaurant survey results reinforce these patterns. For the three overlapping venues (YC, TGB, KCC), reported visit frequencies align closely with the parking survey, indicating consistent behavioral signals across independent instruments. Correlations are high (r≈0.87 for YC and TGB; r≈1.00 for KCC), with the latter largely reflecting strong local-capture effects in Neighborhood H.

Central and more accessible neighborhoods are consistently higher-frequency users in both datasets, while the most isolated neighborhood shows limited engagement outside its local facility. This cross-survey agreement further supports that observed differences reflect actual usage patterns rather than perception bias.

Figure 2 presents calibration curves for the four major facilities, linking perceived problem rates to utilization exceedance probabilities. Across all facilities, the intersection of survey-based PPR with the empirical survival function defines a behavioral threshold that closely aligns with effective capacity.

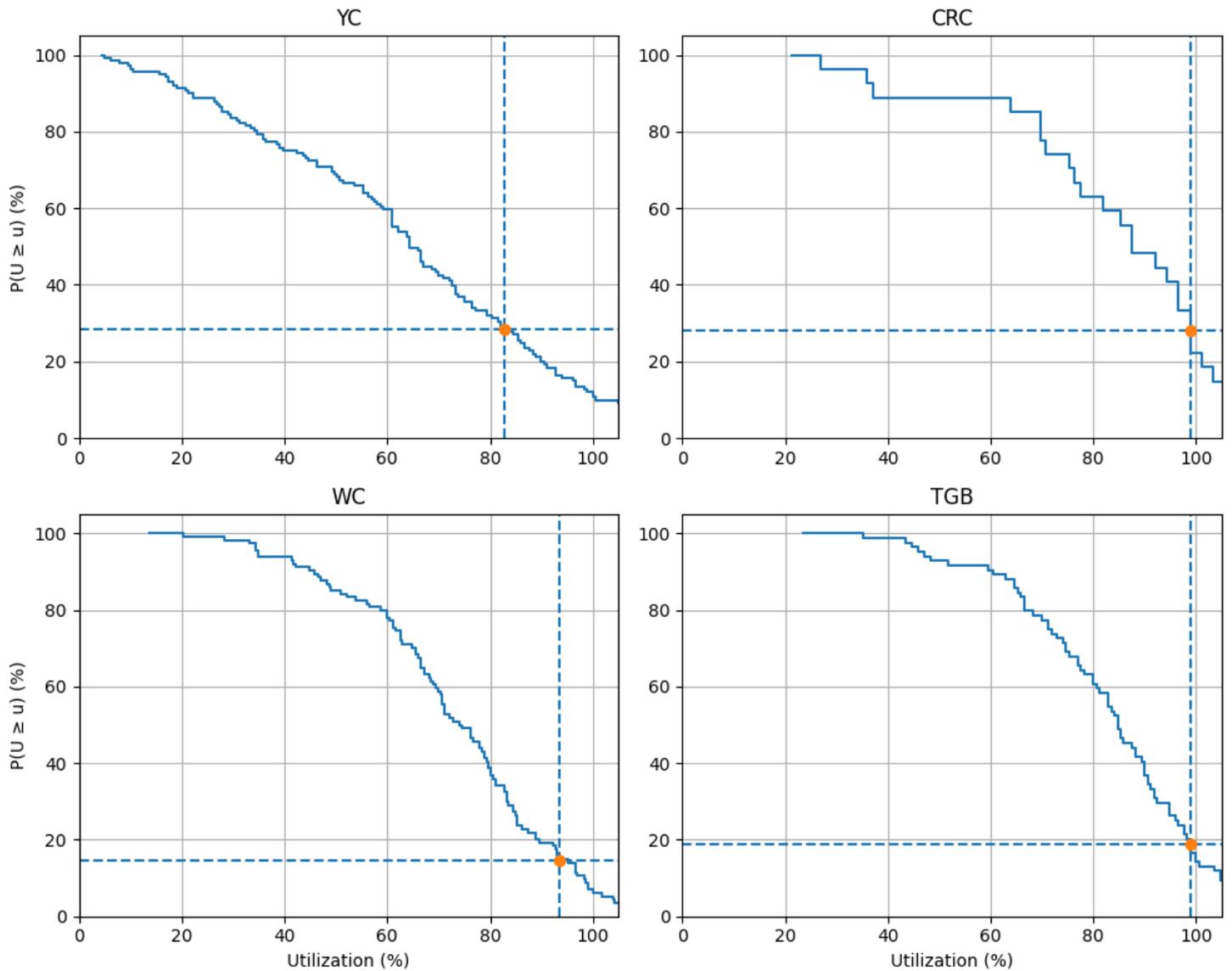

Figure 2 Survey-Calibrated Utilization Thresholds Across Facilities. Solid step curve: Empirical survival function, $P(U \geq u)$. Horizontal dashed line: Survey-based perceived problem rate (PPR). Vertical dashed line: Implied behavioral threshold, $u^*$. Orange point: Calibration point where $P(U \geq u^*) = $ PPR

Table 1 shows that perceived congestion aligns with effective capacity, not nominal capacity. Dissatisfaction emerges at close to nominal capacity for most facilities, for all flat and small parking lots, but much earlier where access is constrained (hilly, less visible lot extensions)

| Facility/System | Survey PPR | Implied threshold (% full) | Full-capacity condition |
|---|---|---|---|
| **YC Parking** | 28.54% | 82.7% | When far-side/ low-visibility spaces are excluded (marina side) |
| **WC Parking** | 14.49% | 93.5% | When far-side/ low-visibility spaces are excluded (behind clinic) |
| **CRC Parking** | 28.08% | 98.9% | ≈ Nominal |
| **TGB Parking** | 18.94% | 98.8% | ≈ Nominal |
| **Meeting Rooms** | 9.3% | ≈100% | ≈ Nominal |
| **Meeting Schedulers** | 40.9% | 25-50% | When required room types and time slots are considered |

Across all facilities, neighborhood differences in perceived problems are highly significant (all p ≪ 0.001), confirming that parking experience is structurally uneven, not random.

Table 1: Survey-Implied Capacity Thresholds and Full-Capacity Conditions

## Discussion

The results show that perceived congestion is driven by effective capacity, not nominal supply. Dissatisfaction emerges when acceptable options are exhausted - spaces that are visible, proximate, and convenient - not when total capacity is fully utilized. This distinction explains why facilities with spatial or informational frictions (e.g., less visible parking areas or constrained room types) reach perceived "full" conditions earlier than more uniform systems.

A central finding is that congestion is best understood as a risk-of-failure perception rather than a response to average conditions. Residents respond not to mean availability, but to the probability of failure - the likelihood that a visit results in an inability to find parking or secure a desired resource. As utilization approaches the behavioral threshold, the variance and uncertainty of outcomes increase, and small changes in conditions lead to disproportionately large increases in perceived difficulty. This response is consistent with loss-averse behavior: unsuccessful experiences weigh more heavily than successful ones, leading to sharp declines in satisfaction near capacity thresholds even when average availability remains adequate.

This interpretation provides a unifying explanation across systems. In parking, failure corresponds to not finding an acceptable space (e.g., close or visible), while in meeting spaces it corresponds to the unavailability of specific room types or time slots. In both cases, nominal capacity overstates functional availability, because users operate within a constrained choice set shaped by spatial, temporal, and informational factors.

The analysis also highlights that perceived congestion is heterogeneous and exposure-dependent. Residents who visit facilities more frequently - or at peak times - experience higher probabilities of failure and therefore report higher problem rates. More importantly, these effects vary systematically across neighborhoods, tenure groups, and usage patterns. Differences in perceived problem rates are statistically significant, indicating that congestion is not uniformly experienced across the community.

This heterogeneity has important implications for measurement. Pulse surveys are effective for tracking overall sentiment and detecting short-term changes, but they primarily capture averages, which can obscure meaningful variation across subgroups. A stable village-wide average may mask high dissatisfaction among specific neighborhoods, newer residents, or high-frequency users. Comprehensive surveys are therefore necessary to collect the explanatory variables - such as location, tenure, age, and usage patterns - needed to segment respondents, identify statistically significant differences, and determine which groups are driving observed outcomes. Without this structure, survey data risk misrepresenting system performance and leading to overly generalized interventions.

The combined calibration framework makes it possible to link these heterogeneous perceptions to observed system behavior. By mapping perceived problem rates to utilization exceedance probabilities, the analysis recovers behavioral thresholds that differ across facilities and implicitly reflect user expectations, constraints, and exposure. These thresholds provide a measurable definition of effective capacity and reveal where operational frictions - rather than physical shortages - are binding.

The practical implication is that many perceived shortages are not structural but operational and informational. Interventions such as demand smoothing, improved visibility of available capacity, better communication, and coordination across facilities can reduce the probability of failure and improve user experience without requiring physical expansion. In systems operating near behavioral thresholds, even small reductions in uncertainty can have disproportionately large effects on perceived service quality.

These findings are consistent with broader evidence that master-planned communities undergo nonlinear changes as they approach buildout [9], where constraints become more binding and system dynamics shift. In such regimes, the probability of failure becomes a more important driver of satisfaction than average utilization, reinforcing the need for measurement frameworks that capture reliability rather than mean conditions alone.

## Conclusion

This study demonstrates that resident surveys, when calibrated with operational data, provide measurable insights into system performance by linking perception to observed behavior. Perceived problems correspond closely to behavioral congestion thresholds, which reflect effective capacity rather than nominal supply. Across parking and meeting spaces, dissatisfaction arises when acceptable options become unreliable - when the probability of failure increases - not when systems are physically full.

Framing congestion as a risk-of-failure perception provides a coherent explanation for observed patterns. This interpretation extends prior utility-based planning frameworks that model resident satisfaction as a trade-off between service quality and cost [8]. While those models capture average conditions, they do not explicitly account for variability and reliability. By incorporating behavioral responses to uncertainty, the present framework aligns more closely with Prospect Theory, in which outcomes are evaluated relative to expectations and losses are weighted more heavily than gains. In this context, service failures - such as the inability to find parking or secure a desired time slot - have a disproportionate impact on perceived satisfaction. This provides a theoretical foundation for why dissatisfaction increases sharply near behavioral capacity thresholds, even when average availability remains acceptable.

Residents respond to uncertainty and variability in outcomes, not to average availability. As utilization approaches the behavioral threshold, the likelihood of unsuccessful experiences increases, and loss-averse responses amplify their impact on satisfaction. This explains why perceived congestion can emerge well before nominal capacity is reached and why small operational improvements can yield large gains in perceived service quality.

The integration of survey and operational data is essential for capturing this dynamic. Pulse surveys provide timely indicators of overall sentiment, but comprehensive surveys are required to identify statistically significant differences across neighborhoods, demographic groups, and usage patterns [7]. These differences reveal how exposure, expectations, and access constraints shape perceived outcomes [3] and enable segmentation of residents into groups with distinct needs. This segmentation supports more precise, targeted interventions and avoids one-size-fits-all solutions.

The key implication is that effective capacity is jointly determined by infrastructure, operations, and user behavior [8]. Improving system performance therefore does not necessarily require expansion. In many cases, reducing the probability of failure - through demand management, improved information, and better coordination - can be more effective and cost-efficient than increasing physical supply [3,5,6,10].

More broadly, the framework developed here provides a unified approach to measuring and managing shared resources in community settings. By combining survey data, continuous operational measurement, and behavioral theory, it improves the interpretability of resident feedback and supports more accurate, actionable decision-making in community planning.